\documentclass[a4paper,twoside]{article}

\usepackage{epsfig}
\usepackage{subcaption}
\usepackage{calc}
\usepackage{amssymb}
\usepackage{amstext}
\usepackage{amsmath}
\usepackage{amsthm}
\usepackage{multicol}
\usepackage{pslatex}
\usepackage{apalike}
\usepackage{algorithm2e}
\usepackage{url}
\usepackage[bottom]{footmisc}
\usepackage{SCITEPRESS}     % Please add other packages that you may need BEFORE the SCITEPRESS.sty package.

\begin{document}

\title{An ETSI GS QKD compliant TLS implementation\thanks{This work has been supported by a government grant managed by the ANR under the Investissement d'avenir program, ref. ANR-17-EURE-004}}

\author{\authorname{Thomas Prévost\sup{1}\orcidAuthor{0009-0000-2224-8574}, Bruno Martin\sup{1}\orcidAuthor{0000-0002-0048-5197} and Olivier Alibart\sup{2}\orcidAuthor{0000-0003-4404-4067}}
\affiliation{\sup{1}I3S, Université Côte d'Azur, CNRS, Sophia-Antipolis, France}
\affiliation{\sup{2}InPhyNi, Université Côte d'Azur, CNRS, Nice, France}
\email{\{thomas.prevost, bruno.martin, olivier.alibart\}@univ-cotedazur.fr}
}


\abstract{A modification of the TLS protocol is presented, using our implementation of the Quantum Key Distribution (QKD) standard ETSI GS QKD 014 v1.1.1. We rely on the Rustls library for this. The TLS protocol is modified while maintaining backward compatibility on the client and server side. We thus wish to participate in the effort to generalize the use of QKD on the Internet. We used our protocol for a video conference call encrypted by QKD. Finally, we analyze the performance of our protocol, comparing the time needed to establish a handshake to that of TLS 1.3.}

\onecolumn \maketitle \normalsize \setcounter{footnote}{0} \vfill

\section{\uppercase{Introduction}}

Quantum computers threaten current public key cryptosystems like RSA and ECC, which are expected to be broken once such machines are operational~\cite{bhatia2020efficient}. This has prompted concerns about “\textit{harvest now, decrypt later}” attacks, where adversaries store encrypted data to decrypt in the future~\cite{paul2022transition}.

Post-quantum cryptography offers alternatives based on quantum-resistant problems, but new attacks continue to emerge~\cite{kaluderovic2022attacks}, raising doubts about their long-term viability. While PKC is still standard for key exchange, we propose replacing it with Quantum Key Distribution (QKD).

QKD enables theoretically perfect forward secrecy by using quantum principles—specifically the no-cloning theorem— to detect eavesdropping in real time~\cite{zygelman2018no}. Keys are exchanged using qubits (typically single photons), and any interception alters their state, alerting the participants. Authentication remains reliant on classical PKC.

Though offering strong security guarantees, QKD faces practical challenges. Device imperfections may allow attacks~\cite{huang2019laser}, and the need for dedicated infrastructure limits its scalability. It's best suited for high-security environments like inter-datacenter links or governmental networks.

Due to fiber loss and the no-cloning theorem, QKD is limited to a few hundred kilometers~\cite{huttner2022long}. Multipath QKD protocols address this with trusted intermediaries~\cite{liu2024multi,icissp25}. ETSI GS QKD 014 v1.1.1 defines a standard interface for managing QKD keys~\cite{etsi2019014}, which we previously verified with ProVerif under specific assumptions~\cite{etsipreviousformalanalysis}.

We present a practical implementation of this standard by integrating QKD into TLS. Our “\emph{TLS-QKD}” protocol replaces the handshake’s public key exchange with a request to a local QKD manager, secured via HTTPS with bilateral authentication. The ETSI standard assumes local networks can safely use classical public key cryptosystems. Once a quantum key is received, symmetric encryption ensures message confidentiality.

TLS-QKD is fully backward compatible: it can interoperate with standard TLS clients and servers. We designed the implementation to align with our formal proof and developed supporting QKD key management tools, including a video conferencing demo using TLS-QKD.

\subsubsection*{Related work}

The use of pre-shared keys (PSK) in TLS is already described in RFCs 9257 and 9258~\cite{housley2022rfc,benjamin2022rfc}. However, we do not rely on this standard, but on a custom protocol that has been formally verified~\cite{etsipreviousformalanalysis}..

\cite{tankovic2024performance} tested an implementation of the ETSI GS QKD 014 standard in real conditions, to measure latency and transmission rate, as well as a complete analysis of the standard's performance in a multi-user environment~\cite{tankovic2024performance2}, and a simulation of the protocol's use cases, with an estimation of possible key rates depending on the location of the sites~\cite{dervisevic2024simulations}.

\cite{martin2024madqci} also proposes an implementation of the ETSI standard on the Madrid QKD network. As many countries develop their QKD network independently, we cannot exhaustively cite all the implementations of the ETSI standard. The final goal would be to set up large QKD networks, for example on the scale of the European continent~\cite{martin2024towards}.

Interestingly, the authors of~\cite{buruaga2025quantum}, who wrote their paper at the same time as us, propose a modification of the TLS protocol based on the ETSI GS QKD 004 standard\footnote{\url{https://www.etsi.org/deliver/etsi_gs/QKD/001_099/004/02.01.01_60/gs_qkd004v020101p.pdf}}.

\hfill

The paper is organized as follow: section~\ref{etsi_pres} introduces the ETSI standard proposal for QKD. Section~\ref{our_implem} presents our implementation, explaining the operation mode of our protocol and the changes to TLS. We evaluate the performance tests performed on our protocol, measuring the time for the handshake with TLS-QKD, compared to that of TLS 1.3. Finally we suggest avenues for improving and we discuss about a possible future for Quantum Key Distribution protocols.

\section{\uppercase{ETSI GS QKD 014 v1.1.1}} \label{etsi_pres}

In this section, we briefly recall the operative mode of the ETSI GS QKD 014 v1.1.1 standard proposal which is mainly focused around a REST interface, through which the different actors interact. The standard defines two types of communication: communication within ``secure zones" (e.g. inside a LAN), where PKC is allowed, and outside communication (e.g. over a WAN), where communications require QKD.

Two types of actors interact in the this protocol: \textbf{KME}: Key Management Entities managing keys within the LAN and exchanging keys with remote KMEs using QKD, and \textbf{SAE}: Secure Application Entities, applications that request keys to KMEs for communication.

SAEs make requests to their datacenter's KME via a REST API, secured by HTTPS. The KME is therefore authenticated by the server certificate. To authenticate itself, the SAE presents a client TLS certificate. This certificate also uniquely identifies the SAE.

\begin{figure}
\includegraphics[width=0.45\textwidth]{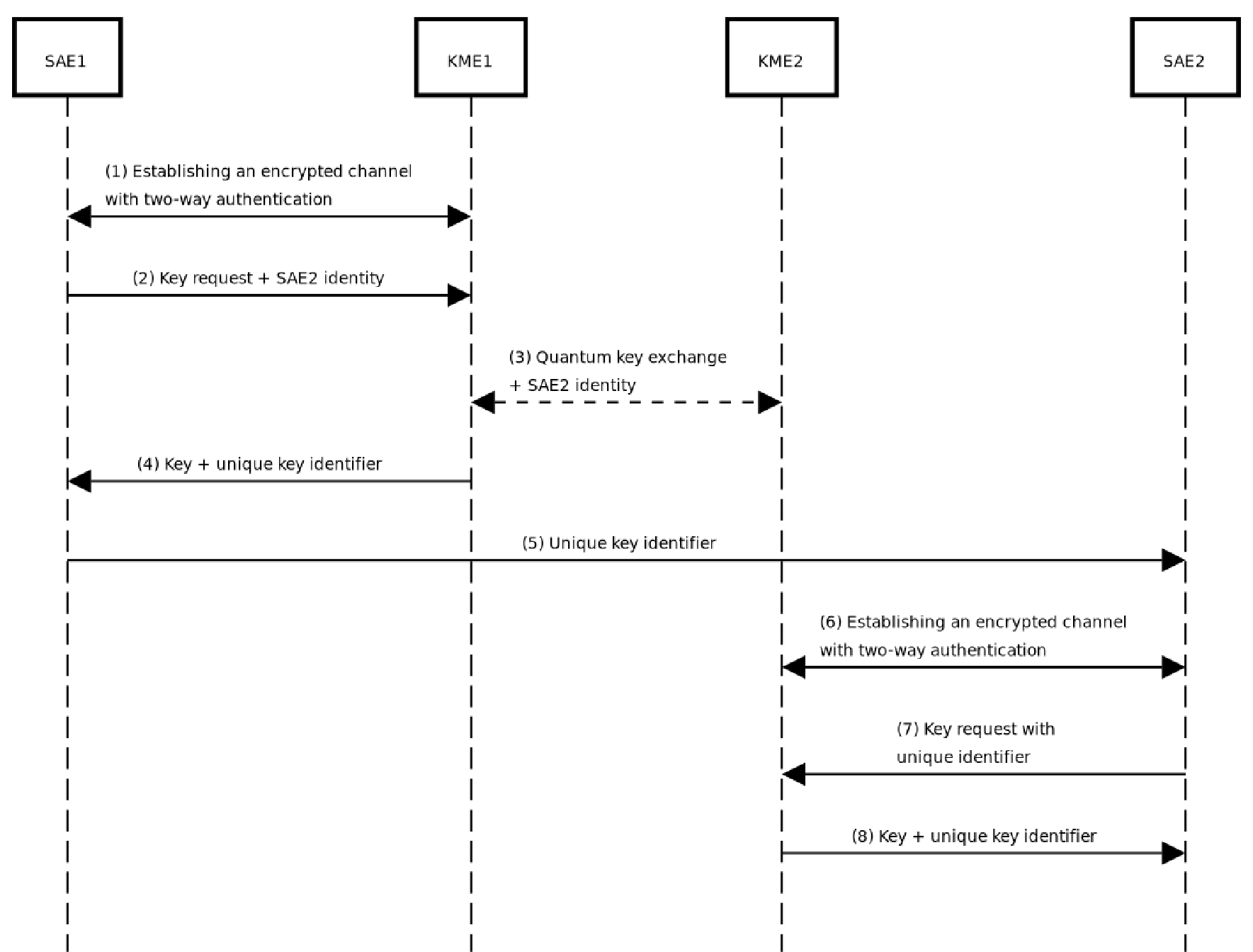}
\caption{This diagram shows a typical quantum key exchange between the initiator SAE~1 (``master") and the SAE~2 "slave", as defined in the standard proposal. SAE~1 makes an authenticated key request to its local KME (KME~1), which communicates the key enciphered within a TLS response to the remote KME (KME~2). SAE~1 then transfers the key identifier to its SAE~2 peer, which requests the key from its local KME.
 } \label{etsi_qkd_protocol}
\end{figure}

Each actor, KME and SAE, is identified by a unique identifier in the network. The keys are identified by their UUID fingerprint, which is also supposed to be unique. Fig.~\ref{etsi_qkd_protocol} shows a key exchange between two SAEs on remote data centers.

The ETSI standard defines the interface and the order of communications. It does not go into cryptographic details, e.g. how QKD is performed or key identifiers are transmitted are considered ``\textit{outside the scope of the document}".

\begin{figure}
\centering
\includegraphics[width=0.4\textwidth]{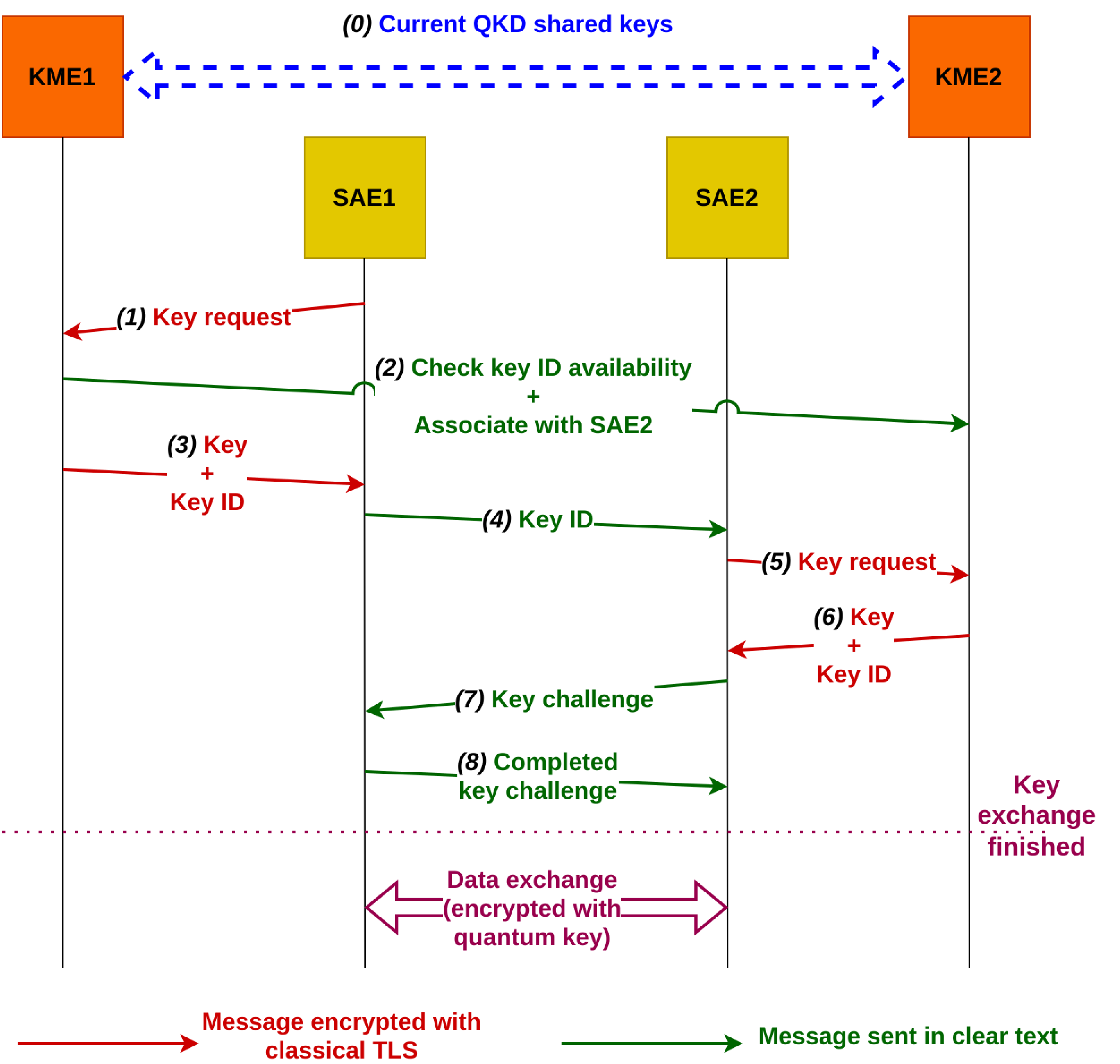}
\caption{Flow of a key exchange using the verified implementation of the ETSI protocol.
 } \label{realistic_etsi_protocol_flow}
\end{figure}

When we formally verified the standard using ProVerif~\cite{etsipreviousformalanalysis},
we determined that the ETSI prototype standard guaranteed the confidentiality and authenticity of the key with the following constraints on the implementation:
\begin{itemize}
    \item The connection between the two KMEs must be authenticated (this is already a prerequisite for the operation of QKD).
    \item The second SAE (``slave") must send a cryptographic challenge to the initiating SAE (``master") to authenticate the latter by ensuring that it owns the correct quantum key.
\end{itemize}

Fig.~\ref{realistic_etsi_protocol_flow} shows an example protocol following an implementation compatible with these security requirements.

\section{\uppercase{Our implementation}} \label{our_implem}

Let us present our Key Management software, as well as our implementation of the
 custom Rustls library, and our video conferencing software based on TLS-QKD.

\subsection{KME key manager}

Key Management Entities (KMEs) are responsible for managing keys within the data center and exchanging keys with their remote counterparts by QKD. For our experiments, we used BBM92~\cite{bennett1992quantum}, based on the entanglement of photons pairs. Any other QKD protocol, however, would have given similar results. 
%(other protocols might have a different maximum geographic distance between remote KMEs). unnecessary
After receiving all the photons, the two remote KMEs share a sequence of random bits. It is necessary to add a Privacy Amplification (PA) step on both sides in order to extract the maximum entropy from these shared bits~\cite{bennett1992experimental}. This step ensures the uniformity of the random bits of the symmetric key, and therefore its security. At the end of the protocol, the protagonists share a secret perfectly random bit string.

Our KME software takes as a parameter a folder in which the key files are located after Privacy Amplification and cuts them into sections of 32 bytes (the keys). If new keys are generated during operation of the KME server, the latter will add them to its database. If the KME exchanges keys with several other KMEs, it takes as parameter the folders containing the keys exchanged with each of its counterparts.

SAEs are authenticated with their client TLS certificate, and identified with the certificate serial number. Indeed, the serial number of the certificate is chosen by the Certificate Authority, and can therefore be unique within the secure zone. The KME associates the serial number with a unique identifier on the network, a 64-bit integer.
 
Each KME also has a unique identifier on the network in the form of a 64-bit integer. The addressing of KMEs is independent of that of SAEs, which means that a KME can have the same address as a SAE.
 
The UUID of the keys is generated from their SHA-1 fingerprint.

We added this route to allow an administrator to detect a failure or an attack in QKD.

In order to inform its remote counterpart of the association between a SAE and its key, the KME also uses the REST protocol, encrypted via HTTPS and authenticated on both sides, between the two KMEs. Bilateral authentication between KMEs is done with client and server X.509 certificates. Note that we use a regular HTTPS protocol to notify the KME of the use of a key, to protect against a ``Harvest now, decrypt later" attacker. So we assume that \emph{at the time of the exchange}, the attacker is not able to break the public key encryption.
 
The KME key manager source code can be retrieved at: \url{https://github.com/thomasarmel/qkd_kme_server/}.

\subsection{Our implementation of TLS with QKD keys}

Our implementation of TLS with QKD keys is a modification of the Rustls library. Our version of Rustls is designed to be backward compatible in both directions. Thus, a TLS-QKD client can connect to a classic TLS server, and a TLS-QKD server can receive classical TLS connections. We could then fear that a malicious actor could carry out a ``\emph{downgrade attack}" to force the protagonists to use classic TLS to weaken the protocol.%~\cite{lei2014three}. 
We are fully aware of this vulnerability, and believe this is an acceptable compromise at this time to facilitate adoption of the protocol. However, the user who needs strict QKD protection could easily disable TLS~1.3 backward compatibility.

Our implementation of TLS-QKD is a modification of the TLS~1.3 protocol~\cite{rescorla2018transport}. Here are the changes we made to the protocol:

\subsubsection{Client and server configuration interface.}
The TLS client and server are two SAEs in the ETSI protocol. Client is the initiating (``master") SAE. They must collect the keys from their local KME. The client and the server take the address and port number of the KME interface as parameters. SAEs authenticate with KMEs using client TLS certificates.

\subsubsection{Protocol version.}
TLS messages contain the protocol version number in two bytes. For example, the code associated with
 TLS version 1.3 is 0x0304. For our TLS-QKD implementation, we arbitrarily chose the number
 \textbf{0x0E00}.

\subsubsection{Client request to KME.}
Equipped with its client TLS certificate and the SAE identifier of the TLS server, the client can make a request to the KME of its ``secure zone" to request a key allowing it to communicate with the remote SAE. The remote SAE is identified by its unique identifier, a 64-bit integer. This number is specified by the programmer when establishing the connection with the KME. The KME then returns the key in base64 format as well as the UUID of this key.

The TLS clients will also request their SAE identifier from their KME:\\
{\small\texttt{https://\{KME\_hostname\}/api/v1/sae/info/me}}

\subsubsection{ClientHello extension.}
The TLS client communicates to the server its SAE identifier as well as the UUID of the key via an extension of the ClientHello message. We add to the extension the Initialization Vector (IV) which will subsequently be used for secret key encryption (the IV could be the output of a key derivation function like PBKDF2). Each type of extension is associated with a 2-byte number. We arbitrarily chose \textbf{0xFEA6} for this ClientHello extension. If the TLS server detects this extension in ClientHello, it determines that the client supports TLS-QKD.

\subsubsection{Server request to KME.}
The TLS server having detected that the client wishes to communicate using TLS-QKD, it makes in turn a request to the KME of its ``secure-zone", to ask for the key associated with the UUID and the identifier of the initiating SAE ``master", received in ClientHello. If the response from the KME is positive, the TLS client and server then share a secret key. However, it remains to correctly authenticate the initiating SAE, which will be done later by a cryptographic challenge.

\subsubsection{ServerHello extension.}
In order to authenticate the client, the TLS server must ensure that the latter is in possession of the quantum key. To do this, it will send him a cryptographic challenge, in the form of a 256 bits random token and a 256 bits random seed, encrypted with the quantum key. The TLS client must send back the same token as well as a \textbf{different} random seed, encrypted with the same quantum key. The challenge is inserted as an extension in the ServerHello response. The 2-byte number we arbitrarily chose for this ServerHello extension is \textbf{0xFEA7}. By finding this extension in the ServerHello response, the client will have confirmation that the server supports TLS-QKD.

\subsubsection{Client challenge acknowledgment.}
After having confirmed that the server supports TLS-QKD, the client must now send the cryptographic challenge back to the server in order to authenticate. After decrypting the ServerHello challenge, the client encrypts it again after changing the random seed. It sends the response to the challenge in the form of a new message type, \textbf{ChallengeAck}. TLS provides a one-byte code for each message type. For example, ApplicationData messages have the code 0x17. For ChallengeAck messages, we arbitrarily chose the code \textbf{0x50}. Once the acknowledgment has been verified by the server, both participants directly start the data transfer using the quantum symmetric key.

The client no longer checks the server's TLS certificate, since the TLS-QKD protocol is sufficient to guarantee its authentication provided that the security assumptions on the KMEs are respected, as proven by ProVerif in~\cite{etsipreviousformalanalysis}. It is also no longer necessary to send a Finished message to authenticate the handshake.

\subsubsection{Symmetric encryption.}
For symmetric encryption, we use AES-AEAD. This is not, however, a security necessity, as authentication is already guaranteed by the protocol.
The key size is arbitrarily fixed to 256 bits, a standard to be quantum safe. The key size is hardcoded in our implementation. Keys are never regenerated in this implementation, we are considering this feature in future work.

\subsubsection{Implementation.}
\begin{figure}
\centering
\includegraphics[width=0.4\textwidth]{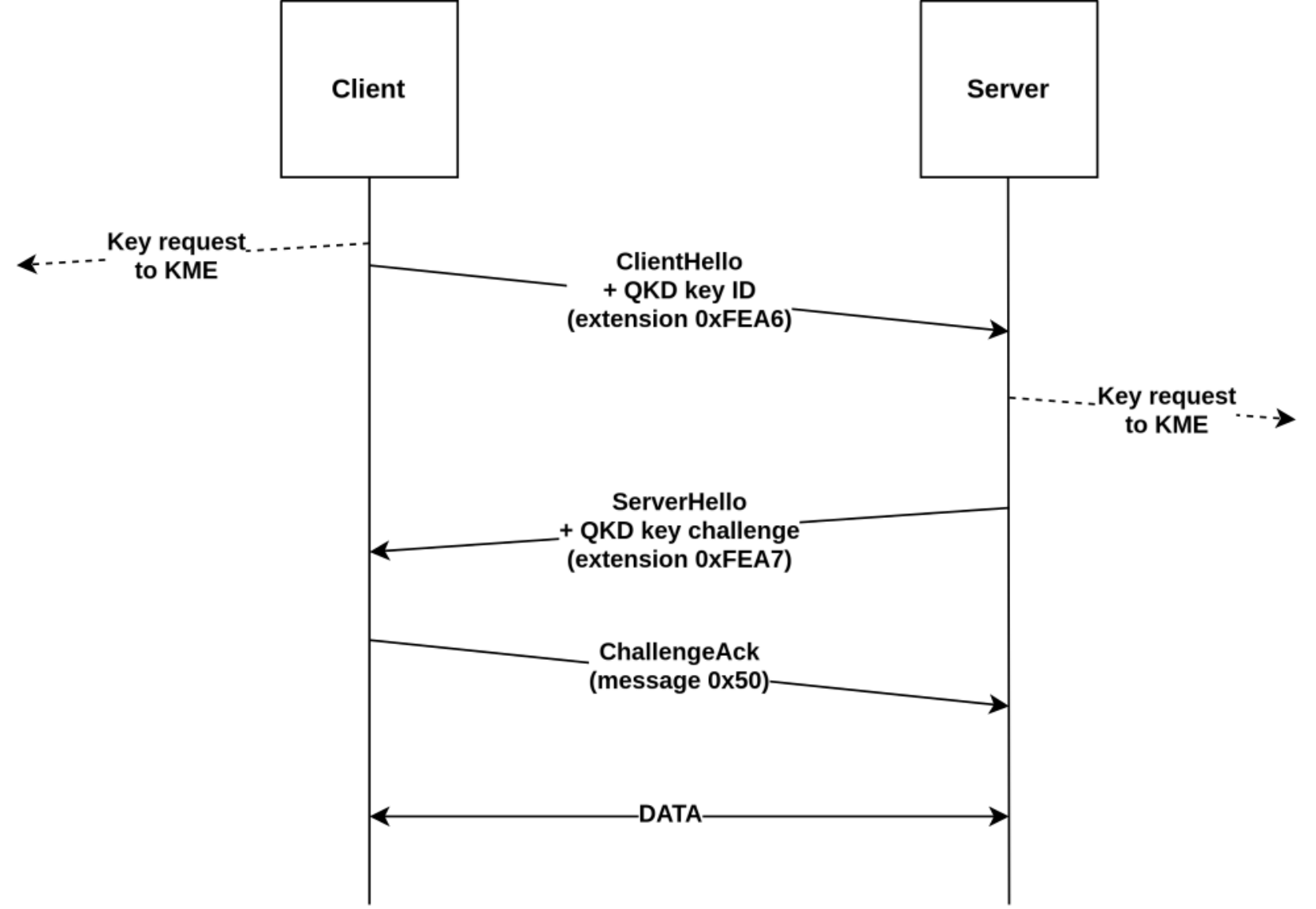}
\caption{Handshake on TLS-QKD} \label{tls-qkd_handshake}
\end{figure}

Fig.~\ref{tls-qkd_handshake} shows a TLS-QKD handshake. Code is available at:~\url{https://github.com/thomasarmel/rustls/tree/qkd}.

\subsection{Proof of concept: a video conference call encrypted by TLS-QKD}

In order to provide a proof of concept of our protocol, we created videoconferencing software encrypted with TLS-QKD. This software is separated into two parts: a server managing the audio and video display, and a client, which captures the video stream from the camera and the audio from the microphone. To make a videoconference call, it is therefore needed to first launch the server on both machines, then the client. Code is available at:~\url{https://github.com/thomasarmel/qkd_camera_streaming_client/}.

During our tests between the INRIA center, in Sophia Antipolis (France) and the InPhyNi site, in Nice (France) at 25km distance, we managed to set up a videoconference with a resolution of 720 pixels and 10 fps. The conversation was absolutely not hampered by sound latency. TLS-QKD typically took less than 1 second to perform the key handshake. %Fig.~\ref{experiment_network} gives the network topology during our experiment.

We also tested the operation of backward compatibility towards classic TLS, the code is available at \url{https://github.com/thomasarmel/rustls/blob/qkd/rustls/tests/qkd.rs}.

\subsection{Protocol performance}

We tested the performance of our protocol on the same sites that were used during the videoconference, 25 kilometers apart. These two sites are connected by a direct optical fiber with a particularly low latency (approximately 1 ms).

We performed 10 measurements of the time required to perform a handshake between the two SAEs with our protocol (on the ``TLS-QKD'' classical link), it takes on average 38.2 ms, with a minimum of 25 ms and a maximum of 66 ms. For comparison, we performed 10 measurements of the latency time of a handshake in TLS 1.3, on the same network: the average delay is 21.5 ms, with a minimum of 17 ms and a maximum of 28 ms. The code used to test the handshake duration is available at \url{https://github.com/thomasarmel/test_https_qkd}.

It should be noted, however, that our network is particularly favorable to our protocol, which requires a lot of message exchanges during the handshake, because the ping latency between the two sites is about 1 ms. A network with higher latency would show a much larger gap in delay between the handshake of our protocol and that of TLS 1.3.

\section{\uppercase{Further improvements}}

The main problem with TLS-QKD is its relative slowness at the handshake, i.e. to exchange the quantum symmetric key. In fact, each SAE must start by establishing a secure connection with its KME.

A total of 29 messages is exchanged between the different actors (SAEs and KMEs) during handshake, including TCP SYN and ACK messages. One way to reduce the number of messages would be to pre-establish a TLS connection between the SAEs and the KMEs. However, this would make our library much less portable, since programs running on SAEs would have to communicate with a background service responsible for keeping the connection with the KME active. Since we are targeting data center use, this compromise could be acceptable.

Another solution would be to rely on the QUIC protocol, which uses UDP instead of TCP. Since SAE-KME communications operate over a LAN, packet loss should not be too much of a problem. This solution would at least reduce latency in communications between SAEs and KMEs.

It would further be possible to pre-establish TLS connections in advance between KMEs. This would work at least as long as the overall network of KMEs is not too large. If the network size becomes too large, we could ensure that only the most frequent inter-KME links pre-establish the TLS connection in advance.

In this implementation, we never regenerate the symmetric key. Adding this feature would increase the security of our protocol. The property of forward secrecy is in fact not assured if an attacker were to discover the quantum key, which is more vulnerable because it is shared between four actors (the two KMEs and the two SAEs).

In our current implementation, the TLS client (initiating SAE) is required to know the TLS server SAE identifier in advance. To avoid managing a directory of correspondence between nodes and their identifiers, we could consider deriving unique identifiers from network addresses. If the IPv6 standard were to be widely adopted, then we could use it as a unique identifier, since IPv6 addresses are the same in the LAN and the WAN.

\section{\uppercase{Conclusion}}

In this paper, we presented a modified TLS protocol which uses keys exchanged by QKD, compliant with the  ETSI GS QKD 014 v1.1.1 standard proposal. Our protocol offers a solution against ``harvest now-decrypt later" attacks.

Our protocol remains vulnerable if the attacker is able to break QKD authentication between KMEs on the fly, since she will be able to carry out a MITM attack. However, this type of scenario seems unlikely today. Indeed, if we still use PKC or PQC for inter-KMEs authentication (for QKD and key requests), it is very unlikely that an attacker would have a quantum computer capable of breaking such a cryptosystem in a short time.

It is backward compatible in both directions with TLS. We have deliberately chosen to leave this backward compatibility despite the risk of ``downgrade attack", in order to facilitate a potential adoption. However, backward compatibility can easily be disabled in the future.

The protocol is based on TLS~1.3, but adds additional configuration for communication with Key Management Entities (KME). The information necessary for the protocol to run is sent in extensions that we added to the ClientHello and ServerHello messages. Additionally, another message is sent by the client at the end of the handshake to confirm their identity, ChallengeAck.

Finally, we showed that our protocol is usable in real application cases, such as videoconferencing. However, the time required for the handshake remains significantly longer than a classic TLS handshake, since many more messages are sent and that the application spends a lot of time waiting for the KMEs stack to return the symmetric keys.

\bibliographystyle{apalike}
{\small
\bibliography{tls_qkd}}

\end{document}